\documentstyle[proceedings,numreferences]{crckapb} 
\newcommand{\stt}{\small\tt}

\begin{opening}
\title{SCALING OF LEVEL STATISTICS\protect\\
       AT THE METAL--INSULATOR TRANSITION}

\author{I.~Kh. Zharekeshev}
\author{B.~Kramer}
\institute{I. Institut f\"ur Theoretische Physik, Universit\"at Hamburg,\\
Jungiusstrasse 9, D--20355 Hamburg, Germany}

\end{opening}

\runningtitle{SCALING OF LEVEL STATISTICS}

\begin{document}

% The \begin{document} command comes after the \end{opening}
% command.

\begin{abstract}
Using the Anderson model for disordered systems 
the fluctuations in electron spectra near the 
metal--insulator transition were numerically calculated 
for lattices of sizes up to $28\times 28\times 28$ sites. 
The results show a finite--size scaling of both the level spacing 
distribution and the variance of number of states in a given
energy interval,
that allows to locate the critical
point and to determine the critical exponent of the localization length.  
\end{abstract}
The statistical description of energy spectra of 
disordered quantum systems 
is based on the random--matrix theory [1--3].
%\cite{RMT,Efetov83,Altshuler86}.
One of the main properties 
of the spectra of random--matrices is the repulsion between 
their eigenvalues. 
For a disordered metal, such a correlation of energy levels is caused 
by a pronounced overlap of delocalized one--electron states.
By increasing the disorder of a random potential the system is known 
to undergo a metal--insulator transition (MIT).
On the insulating side of the MIT, the energy levels 
of the localized states are not correlated due to vanishing
of the level repulsion.
A central question in the problem of level statistics 
is how a character of spectral fluctuations
varies 
when the system changes
from the delocalized to the localized regime.  

On the metallic side 
of the MIT the distribution function $P(s)$ of neighboring spacings between
levels can be described by the Wigner formula\,\cite{Imry87}
\begin{equation}
P_{W}(s)=\frac{\pi}{2}\,s\,\exp \left({-\frac{\pi}{4} s^2}\right),
\label{Wigner}
\end{equation}
where $s$ is measured in units of the mean inter--level spacing $\Delta$.  In the\\ 
\vspace*{2mm}
{\small\centerline {93}\\
\vspace*{0.5mm}
\footnotesize\it
{H.A.Cerdeira et al. (eds.), Quantum Dynamics of Submicron Structures,~93-98\\
\copyright~1995 Kluwer Academic Publishers. Printed in the Netherlands.}}\\
localized regime the spacings are distributed according to a Poisson law
\begin{equation}
P_{P}(s)=\exp (- s).
\label{Poisson}
\end{equation}
The crossover of 
$P(s)$ 
between the Wigner (\ref{Wigner}) and the Poisson statistics (\ref{Poisson})
which accompanies the MIT was extensively investigated both 
numerically [5--8]  
%\cite{Altshuler88,Evangelou92,Shklovskii93,Hofstetter94}
and analytically \cite{Aronov94,Thisproc}.
It was pointed out 
by Shklovskii {\it et al} \cite{Shklovskii93} that 
the level spacing distribution exhibits 
critical behavior near the MIT.  
Therefore we focus on 
the finite--size scaling properties of the distribution $P(s)$.
In addition, we analyze the variance of the number of energy levels
$\langle [\delta N(E)^2]\rangle$ 
as a function of the average level number 
$\langle N(E)\rangle$ 
in a specified energy interval $E$, 
and study its universal peculiarities at the critical point.    

One of the simplest models describing 
a disorder--induced MIT is the Anderson
model
\begin{equation}
H=\sum_{n}\epsilon_{n}^{} a_{n}^{+} a_{n}^{} +
          \sum_{n\neq m} (a_{n}^{+} a_{m}^{} + a_{n}^{} a_{m}^{+}).
\label{Hamiltonian}
\end{equation}
Here
$a_{n}^{+}$ and $a_{n}^{}$ are the creation and annihilation operators of
an electron at a site $n$ in a lattice; 
$m$ denote the nearest neighbors of $n$. 
The on-site energy $\epsilon_{n}$ 
is measured in units of the overlap integral between adjacent sites 
and is uniformly distributed variable in the interval from $-W/2$ to $W/2$.  
The parameter $W$ specifies the degree of the disorder of the system.
The critical disorder
of the MIT which occurs in the middle of the band 
corresponds to $W_{c}\approx 16.5$ \cite{Kramer83}.
For a finite system when $W\ll W_{c}$, the level statistic is close to 
(\ref{Wigner}) and when $W\gg W_{c}$ 
it obeys~(\ref{Poisson})~\cite{Altshuler88}. 
 
In order to find the electron spectrum 
in the critical region 
we diagonalized numerically 
the real symmetric 
Hamiltonian (\ref{Hamiltonian}) 
for simple cubic lattices of the size $L\times L\times L$ 
with periodic boundary conditions. 
The Lanczos algorithm for eigenvalue problem was used in a version that was
especially designed
for very sparse and big matrices with hierarchic structure. 
We applied the algorithm for cubes with 
$L = 6, 8, 12, 16, 20, 24, 28$ at various degree of the disorder 
close to $W_{c}$.
We consider an energy interval which is centered 
at $\epsilon =0$ and has such a width  that it 
contains a half of all eigenvalues.
As we are interested in sample--to--sample fluctuations in the spectra
 the calculations were carried out for
ensembles of different random 
configurations.
After unfolding the spectrum 
the histograms of several spectral distributions 
 were constructed by use of $10^{5}$ spacings calculated  
for each pair of $\{L,W\}$. 

\begin{figure}[htb]
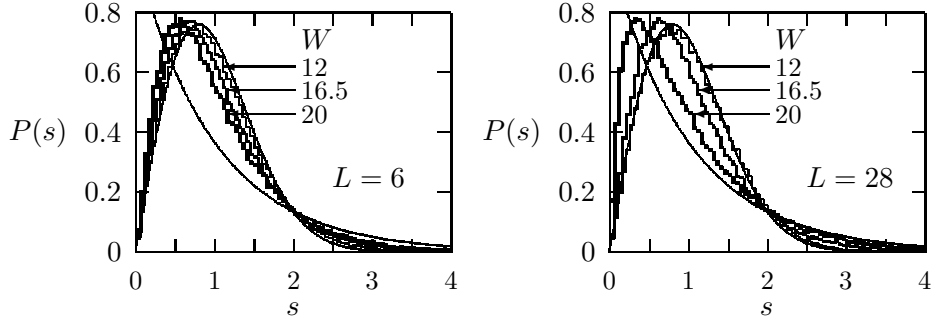

%\vspace{4.4cm}
\unitlength1cm
\begin{minipage}[t]{5.9cm}

}
\end{picture}
\end{minipage}
\caption[]{The level--spacing distribution $P(s)$
for various disorders near the transition.
Continuous curves correspond the
Wigner (\ref{Wigner}) and the Poisson (\ref{Poisson}) 
distributions for the metallic and insulating phases, respectively.}
\label{fig1}
\end{figure}

Fig.~\ref{fig1}\, displays the  distribution function
$P(s)$ calculated near the MIT for two cubes of sizes $L=6$ and $L=28$.
By increasing the disorder $W$ the spacing distribution for both $L$ varies
continuously from $P_{W}(s)$ to $P_{P}(s)$ over all range of spacings. 
Results of calculations for other, intermediate sizes confirm this continuous
crossover.
But the behavior of this crossover substantially depends on the size of the 
cube. 
One can see 
that $P(s)$ changes faster
between 
(\ref{Wigner}) and (\ref{Poisson}) for $L=28$ than for $L=6$.
The size--dependence of $P(s)$ is observed on both sides of the MIT.
However at the transition point when $W_{c}=16.5$ the spacing distribution 
has almost the same form 
for all $L$ from 6 to 28.
The independence of the level spacing distribution 
on the size of system at $W_{c}$ is in good agreement with 
predictions of the new universal level statistics at $L\to \infty $ 
which exists
exactly at the transition\cite{Shklovskii93}. 

In order to study the scaling properties of the spacing distribution
in more detail we introduce the 
quantity 
$\alpha = \int_{0}^{s} (P(s)-P_{W}(s)) ds/\int_{0}^{s} (P_{P}(s)-P_{W}(s)) ds$, 
which describes a normalized deviation of $P(s)$ 
from the Wigner distribution (\ref{Wigner}). 
We chose $s=0.473$, the crossing point of $P_{W}(s)$ and $P_{P}(s)$, 
in order to study $P(s)$ in the range of small spacings. 
In this case $\alpha$ corresponds to a ``strength'' of the  repulsion of
two consecutive levels when the separation between them is less than 
average spacing $\Delta$.
In the thermodynamic limit $\alpha=0$ for $W<W_{c}$, 
and $\alpha=1$ for  $W>W_{c}$. 
For finite $L$ it is reasonable to assume a scaling law
\begin{equation}
\alpha(W,L)=f(L/\xi(W)),
\label{scaling}
\end{equation}
where $\xi(W)$ is the correlation length of the transition.

Fig.~\ref{fig2} shows 
the $L$-dependence of the parameter 
$\alpha$ near the critical 
point. In the insulating regime $\alpha$
grows with $L$ approaching its limit $\alpha=1$, whereas in the metallic
phase $\alpha$ decreases to zero.  The change of the sign of the size effect
takes place at the critical point.
Therefore we can determine very
accurately 
the critical value of the disorder $W_{c}=16.35\pm 0.15$. 
Similar results for smaller systems  were earlier 
obtained in Ref. \cite{Shklovskii93,Hofstetter94} for 
$s_{0}=2$ which corresponds to using $P(s)$ only in the range 
of asymptotically large spacings. 
In order to obtain the scaling curve (\ref{scaling}), 
on  which all data points collapse, we shifted  $\ln L$ by $\ln \xi(W)$
separately for each $W$.
The overlap between adjacent values of $W$ allows to fit  
most of the points onto two branches for $W<W_{c}$ and
$W>W_{c}$ corresponding to localized and delocalized regimes, 
respectively (Fig.~\ref{fig3}). As s result one finds 
the disorder dependence of the correlation length $\xi(W)$.
\begin{figure}[tb]
%\vspace{4.9cm}
\unitlength1cm
\begin{minipage}[t]{6.3cm}
\begin{picture}(6.3,4.7)
\put(-0.6,-0.2){
%%\input{plot3}
%%------------------------------------------------------
% GNUPLOT: LaTeX picture
\setlength{\unitlength}{0.240900pt}
\ifx\plotpoint\undefined\newsavebox{\plotpoint}\fi
\sbox{\plotpoint}{\rule[-0.200pt]{0.400pt}{0.400pt}}%
\begin{picture}(840,566)(0,0)
\font\gnuplot=cmr10 at 10pt
\gnuplot
\sbox{\plotpoint}{\rule[-0.200pt]{0.400pt}{0.400pt}}%
\put(220.0,113.0){\rule[-0.200pt]{4.818pt}{0.400pt}}
\put(198,113){\makebox(0,0)[r]{$0.01$}}
\put(756.0,113.0){\rule[-0.200pt]{4.818pt}{0.400pt}}
\put(220.0,169.0){\rule[-0.200pt]{2.409pt}{0.400pt}}
\put(766.0,169.0){\rule[-0.200pt]{2.409pt}{0.400pt}}
\put(220.0,202.0){\rule[-0.200pt]{2.409pt}{0.400pt}}
\put(766.0,202.0){\rule[-0.200pt]{2.409pt}{0.400pt}}
\put(220.0,226.0){\rule[-0.200pt]{2.409pt}{0.400pt}}
\put(766.0,226.0){\rule[-0.200pt]{2.409pt}{0.400pt}}
\put(220.0,244.0){\rule[-0.200pt]{2.409pt}{0.400pt}}
\put(766.0,244.0){\rule[-0.200pt]{2.409pt}{0.400pt}}
\put(220.0,258.0){\rule[-0.200pt]{2.409pt}{0.400pt}}
\put(766.0,258.0){\rule[-0.200pt]{2.409pt}{0.400pt}}
\put(220.0,271.0){\rule[-0.200pt]{2.409pt}{0.400pt}}
\put(766.0,271.0){\rule[-0.200pt]{2.409pt}{0.400pt}}
\put(220.0,282.0){\rule[-0.200pt]{2.409pt}{0.400pt}}
\put(766.0,282.0){\rule[-0.200pt]{2.409pt}{0.400pt}}
\put(220.0,291.0){\rule[-0.200pt]{2.409pt}{0.400pt}}
\put(766.0,291.0){\rule[-0.200pt]{2.409pt}{0.400pt}}
\put(220.0,300.0){\rule[-0.200pt]{4.818pt}{0.400pt}}
\put(198,300){\makebox(0,0)[r]{$0.1$}}
\put(756.0,300.0){\rule[-0.200pt]{4.818pt}{0.400pt}}
\put(220.0,356.0){\rule[-0.200pt]{2.409pt}{0.400pt}}
\put(766.0,356.0){\rule[-0.200pt]{2.409pt}{0.400pt}}
\put(220.0,389.0){\rule[-0.200pt]{2.409pt}{0.400pt}}
\put(766.0,389.0){\rule[-0.200pt]{2.409pt}{0.400pt}}
\put(220.0,412.0){\rule[-0.200pt]{2.409pt}{0.400pt}}
\put(766.0,412.0){\rule[-0.200pt]{2.409pt}{0.400pt}}
\put(220.0,430.0){\rule[-0.200pt]{2.409pt}{0.400pt}}
\put(766.0,430.0){\rule[-0.200pt]{2.409pt}{0.400pt}}
\put(220.0,445.0){\rule[-0.200pt]{2.409pt}{0.400pt}}
\put(766.0,445.0){\rule[-0.200pt]{2.409pt}{0.400pt}}
\put(220.0,458.0){\rule[-0.200pt]{2.409pt}{0.400pt}}
\put(766.0,458.0){\rule[-0.200pt]{2.409pt}{0.400pt}}
\put(220.0,469.0){\rule[-0.200pt]{2.409pt}{0.400pt}}
\put(766.0,469.0){\rule[-0.200pt]{2.409pt}{0.400pt}}
\put(220.0,478.0){\rule[-0.200pt]{2.409pt}{0.400pt}}
\put(766.0,478.0){\rule[-0.200pt]{2.409pt}{0.400pt}}
\put(220.0,487.0){\rule[-0.200pt]{4.818pt}{0.400pt}}
\put(198,487){\makebox(0,0)[r]{$1$}}
\put(756.0,487.0){\rule[-0.200pt]{4.818pt}{0.400pt}}
\put(220.0,543.0){\rule[-0.200pt]{2.409pt}{0.400pt}}
\put(766.0,543.0){\rule[-0.200pt]{2.409pt}{0.400pt}}
\put(220.0,113.0){\rule[-0.200pt]{0.400pt}{4.818pt}}
\put(220,68){\makebox(0,0){4}}
\put(220.0,523.0){\rule[-0.200pt]{0.400pt}{4.818pt}}
\put(285.0,113.0){\rule[-0.200pt]{0.400pt}{4.818pt}}
\put(285.0,523.0){\rule[-0.200pt]{0.400pt}{4.818pt}}
\put(351.0,113.0){\rule[-0.200pt]{0.400pt}{4.818pt}}
\put(351,68){\makebox(0,0){12}}
\put(351.0,523.0){\rule[-0.200pt]{0.400pt}{4.818pt}}
\put(416.0,113.0){\rule[-0.200pt]{0.400pt}{4.818pt}}
\put(416.0,523.0){\rule[-0.200pt]{0.400pt}{4.818pt}}
\put(482.0,113.0){\rule[-0.200pt]{0.400pt}{4.818pt}}
\put(482,68){\makebox(0,0){20}}
\put(482.0,523.0){\rule[-0.200pt]{0.400pt}{4.818pt}}
\put(547.0,113.0){\rule[-0.200pt]{0.400pt}{4.818pt}}
\put(547.0,523.0){\rule[-0.200pt]{0.400pt}{4.818pt}}
\put(612.0,113.0){\rule[-0.200pt]{0.400pt}{4.818pt}}
\put(612,68){\makebox(0,0){28}}
\put(612.0,523.0){\rule[-0.200pt]{0.400pt}{4.818pt}}
\put(678.0,113.0){\rule[-0.200pt]{0.400pt}{4.818pt}}
\put(678.0,523.0){\rule[-0.200pt]{0.400pt}{4.818pt}}
\put(743.0,113.0){\rule[-0.200pt]{0.400pt}{4.818pt}}
\put(743,68){\makebox(0,0){36}}
\put(743.0,523.0){\rule[-0.200pt]{0.400pt}{4.818pt}}
\put(220.0,113.0){\rule[-0.200pt]{133.940pt}{0.400pt}}
\put(776.0,113.0){\rule[-0.200pt]{0.400pt}{103.587pt}}
\put(220.0,543.0){\rule[-0.200pt]{133.940pt}{0.400pt}}
\put(133,373){\makebox(0,0){$\alpha(L)$}}
\put(498,23){\makebox(0,0){$L$}}
\put(760,219){\makebox(0,0)[r]{\small 12}}
\put(760,285){\makebox(0,0)[r]{\small 14}}
\put(760,325){\makebox(0,0)[r]{\small 15}}
\put(760,367){\makebox(0,0)[r]{\small 16}}
\put(711,385){\makebox(0,0)[r]{\small 16.5}}
\put(760,401){\makebox(0,0)[r]{\small 17}}
\put(760,430){\makebox(0,0)[r]{\small 18}}
\put(760,466){\makebox(0,0)[r]{\small 20}}
\put(760,508){\makebox(0,0)[r]{\small $W$}}
\put(220.0,113.0){\rule[-0.200pt]{0.400pt}{103.587pt}}
\put(253,293){\raisebox{-.8pt}{\makebox(0,0){$\Diamond$}}}
\put(285,290){\raisebox{-.8pt}{\makebox(0,0){$\Diamond$}}}
\put(351,273){\raisebox{-.8pt}{\makebox(0,0){$\Diamond$}}}
\put(416,245){\raisebox{-.8pt}{\makebox(0,0){$\Diamond$}}}
\put(482,229){\raisebox{-.8pt}{\makebox(0,0){$\Diamond$}}}
\put(547,217){\raisebox{-.8pt}{\makebox(0,0){$\Diamond$}}}
\put(612,217){\raisebox{-.8pt}{\makebox(0,0){$\Diamond$}}}
\put(253,333){\makebox(0,0){$+$}}
\put(285,331){\makebox(0,0){$+$}}
\put(351,318){\makebox(0,0){$+$}}
\put(416,308){\makebox(0,0){$+$}}
\put(482,302){\makebox(0,0){$+$}}
\put(547,299){\makebox(0,0){$+$}}
\put(612,285){\makebox(0,0){$+$}}
\put(253,356){\raisebox{-.8pt}{\makebox(0,0){$\Box$}}}
\put(285,354){\raisebox{-.8pt}{\makebox(0,0){$\Box$}}}
\put(351,345){\raisebox{-.8pt}{\makebox(0,0){$\Box$}}}
\put(416,337){\raisebox{-.8pt}{\makebox(0,0){$\Box$}}}
\put(482,329){\raisebox{-.8pt}{\makebox(0,0){$\Box$}}}
\put(547,329){\raisebox{-.8pt}{\makebox(0,0){$\Box$}}}
\put(612,325){\raisebox{-.8pt}{\makebox(0,0){$\Box$}}}
\put(253,369){\makebox(0,0){$\times$}}
\put(285,371){\makebox(0,0){$\times$}}
\put(351,372){\makebox(0,0){$\times$}}
\put(416,368){\makebox(0,0){$\times$}}
\put(482,367){\makebox(0,0){$\times$}}
\put(547,365){\makebox(0,0){$\times$}}
\put(612,367){\makebox(0,0){$\times$}}
\put(253,377){\makebox(0,0){$\triangle$}}
\put(285,383){\makebox(0,0){$\triangle$}}
\put(351,383){\makebox(0,0){$\triangle$}}
\put(416,384){\makebox(0,0){$\triangle$}}
\put(482,383){\makebox(0,0){$\triangle$}}
\put(547,384){\makebox(0,0){$\triangle$}}
\put(612,385){\makebox(0,0){$\triangle$}}
\put(253,390){\makebox(0,0){$\star$}}
\put(285,392){\makebox(0,0){$\star$}}
\put(351,393){\makebox(0,0){$\star$}}
\put(416,397){\makebox(0,0){$\star$}}
\put(482,400){\makebox(0,0){$\star$}}
\put(547,401){\makebox(0,0){$\star$}}
\put(612,401){\makebox(0,0){$\star$}}
\put(253,398){\circle{18}}
\put(285,405){\circle{18}}
\put(351,412){\circle{18}}
\put(416,416){\circle{18}}
\put(482,423){\circle{18}}
\put(547,426){\circle{18}}
\put(612,430){\circle{18}}
\put(253,421){\circle*{18}}
\put(285,427){\circle*{18}}
\put(351,439){\circle*{18}}
\put(416,447){\circle*{18}}
\put(482,454){\circle*{18}}
\put(547,456){\circle*{18}}
\put(612,461){\circle*{18}}
\end{picture}
}
\end{picture}
\caption{The quantity $\alpha$
as a function of the size $L$ 
at different degrees of disorder near the metal--insulator transition.}
\label{fig2}
\end{minipage}
\hfill
\begin{minipage}[t]{5.6cm}
\begin{picture}(5.6,4.7)
\put(-1.0,-0.2){
%%\input{plot4}
%%------------------------------------------------------
% GNUPLOT: LaTeX picture
\setlength{\unitlength}{0.240900pt}
\ifx\plotpoint\undefined\newsavebox{\plotpoint}\fi
\sbox{\plotpoint}{\rule[-0.200pt]{0.400pt}{0.400pt}}%
\begin{picture}(825,566)(0,0)
\font\gnuplot=cmr10 at 10pt
\gnuplot
\sbox{\plotpoint}{\rule[-0.200pt]{0.400pt}{0.400pt}}%
\put(220.0,113.0){\rule[-0.200pt]{4.818pt}{0.400pt}}
\put(198,113){\makebox(0,0)[r]{$0.01$}}
\put(741.0,113.0){\rule[-0.200pt]{4.818pt}{0.400pt}}
\put(220.0,169.0){\rule[-0.200pt]{2.409pt}{0.400pt}}
\put(751.0,169.0){\rule[-0.200pt]{2.409pt}{0.400pt}}
\put(220.0,202.0){\rule[-0.200pt]{2.409pt}{0.400pt}}
\put(751.0,202.0){\rule[-0.200pt]{2.409pt}{0.400pt}}
\put(220.0,226.0){\rule[-0.200pt]{2.409pt}{0.400pt}}
\put(751.0,226.0){\rule[-0.200pt]{2.409pt}{0.400pt}}
\put(220.0,244.0){\rule[-0.200pt]{2.409pt}{0.400pt}}
\put(751.0,244.0){\rule[-0.200pt]{2.409pt}{0.400pt}}
\put(220.0,258.0){\rule[-0.200pt]{2.409pt}{0.400pt}}
\put(751.0,258.0){\rule[-0.200pt]{2.409pt}{0.400pt}}
\put(220.0,271.0){\rule[-0.200pt]{2.409pt}{0.400pt}}
\put(751.0,271.0){\rule[-0.200pt]{2.409pt}{0.400pt}}
\put(220.0,282.0){\rule[-0.200pt]{2.409pt}{0.400pt}}
\put(751.0,282.0){\rule[-0.200pt]{2.409pt}{0.400pt}}
\put(220.0,291.0){\rule[-0.200pt]{2.409pt}{0.400pt}}
\put(751.0,291.0){\rule[-0.200pt]{2.409pt}{0.400pt}}
\put(220.0,300.0){\rule[-0.200pt]{4.818pt}{0.400pt}}
\put(198,300){\makebox(0,0)[r]{$0.1$}}
\put(741.0,300.0){\rule[-0.200pt]{4.818pt}{0.400pt}}
\put(220.0,356.0){\rule[-0.200pt]{2.409pt}{0.400pt}}
\put(751.0,356.0){\rule[-0.200pt]{2.409pt}{0.400pt}}
\put(220.0,389.0){\rule[-0.200pt]{2.409pt}{0.400pt}}
\put(751.0,389.0){\rule[-0.200pt]{2.409pt}{0.400pt}}
\put(220.0,412.0){\rule[-0.200pt]{2.409pt}{0.400pt}}
\put(751.0,412.0){\rule[-0.200pt]{2.409pt}{0.400pt}}
\put(220.0,430.0){\rule[-0.200pt]{2.409pt}{0.400pt}}
\put(751.0,430.0){\rule[-0.200pt]{2.409pt}{0.400pt}}
\put(220.0,445.0){\rule[-0.200pt]{2.409pt}{0.400pt}}
\put(751.0,445.0){\rule[-0.200pt]{2.409pt}{0.400pt}}
\put(220.0,458.0){\rule[-0.200pt]{2.409pt}{0.400pt}}
\put(751.0,458.0){\rule[-0.200pt]{2.409pt}{0.400pt}}
\put(220.0,469.0){\rule[-0.200pt]{2.409pt}{0.400pt}}
\put(751.0,469.0){\rule[-0.200pt]{2.409pt}{0.400pt}}
\put(220.0,478.0){\rule[-0.200pt]{2.409pt}{0.400pt}}
\put(751.0,478.0){\rule[-0.200pt]{2.409pt}{0.400pt}}
\put(220.0,487.0){\rule[-0.200pt]{4.818pt}{0.400pt}}
\put(198,487){\makebox(0,0)[r]{$1$}}
\put(741.0,487.0){\rule[-0.200pt]{4.818pt}{0.400pt}}
\put(220.0,543.0){\rule[-0.200pt]{2.409pt}{0.400pt}}
\put(751.0,543.0){\rule[-0.200pt]{2.409pt}{0.400pt}}
\put(220.0,113.0){\rule[-0.200pt]{0.400pt}{2.409pt}}
\put(220.0,533.0){\rule[-0.200pt]{0.400pt}{2.409pt}}
\put(244.0,113.0){\rule[-0.200pt]{0.400pt}{2.409pt}}
\put(244.0,533.0){\rule[-0.200pt]{0.400pt}{2.409pt}}
\put(261.0,113.0){\rule[-0.200pt]{0.400pt}{2.409pt}}
\put(261.0,533.0){\rule[-0.200pt]{0.400pt}{2.409pt}}
\put(274.0,113.0){\rule[-0.200pt]{0.400pt}{2.409pt}}
\put(274.0,533.0){\rule[-0.200pt]{0.400pt}{2.409pt}}
\put(285.0,113.0){\rule[-0.200pt]{0.400pt}{2.409pt}}
\put(285.0,533.0){\rule[-0.200pt]{0.400pt}{2.409pt}}
\put(294.0,113.0){\rule[-0.200pt]{0.400pt}{2.409pt}}
\put(294.0,533.0){\rule[-0.200pt]{0.400pt}{2.409pt}}
\put(301.0,113.0){\rule[-0.200pt]{0.400pt}{2.409pt}}
\put(301.0,533.0){\rule[-0.200pt]{0.400pt}{2.409pt}}
\put(308.0,113.0){\rule[-0.200pt]{0.400pt}{2.409pt}}
\put(308.0,533.0){\rule[-0.200pt]{0.400pt}{2.409pt}}
\put(315.0,113.0){\rule[-0.200pt]{0.400pt}{4.818pt}}
\put(315,68){\makebox(0,0){$0.01$}}
\put(315.0,523.0){\rule[-0.200pt]{0.400pt}{4.818pt}}
\put(355.0,113.0){\rule[-0.200pt]{0.400pt}{2.409pt}}
\put(355.0,533.0){\rule[-0.200pt]{0.400pt}{2.409pt}}
\put(379.0,113.0){\rule[-0.200pt]{0.400pt}{2.409pt}}
\put(379.0,533.0){\rule[-0.200pt]{0.400pt}{2.409pt}}
\put(396.0,113.0){\rule[-0.200pt]{0.400pt}{2.409pt}}
\put(396.0,533.0){\rule[-0.200pt]{0.400pt}{2.409pt}}
\put(409.0,113.0){\rule[-0.200pt]{0.400pt}{2.409pt}}
\put(409.0,533.0){\rule[-0.200pt]{0.400pt}{2.409pt}}
\put(420.0,113.0){\rule[-0.200pt]{0.400pt}{2.409pt}}
\put(420.0,533.0){\rule[-0.200pt]{0.400pt}{2.409pt}}
\put(429.0,113.0){\rule[-0.200pt]{0.400pt}{2.409pt}}
\put(429.0,533.0){\rule[-0.200pt]{0.400pt}{2.409pt}}
\put(437.0,113.0){\rule[-0.200pt]{0.400pt}{2.409pt}}
\put(437.0,533.0){\rule[-0.200pt]{0.400pt}{2.409pt}}
\put(444.0,113.0){\rule[-0.200pt]{0.400pt}{2.409pt}}
\put(444.0,533.0){\rule[-0.200pt]{0.400pt}{2.409pt}}
\put(450.0,113.0){\rule[-0.200pt]{0.400pt}{4.818pt}}
\put(450,68){\makebox(0,0){$0.1$}}
\put(450.0,523.0){\rule[-0.200pt]{0.400pt}{4.818pt}}
\put(491.0,113.0){\rule[-0.200pt]{0.400pt}{2.409pt}}
\put(491.0,533.0){\rule[-0.200pt]{0.400pt}{2.409pt}}
\put(514.0,113.0){\rule[-0.200pt]{0.400pt}{2.409pt}}
\put(514.0,533.0){\rule[-0.200pt]{0.400pt}{2.409pt}}
\put(531.0,113.0){\rule[-0.200pt]{0.400pt}{2.409pt}}
\put(531.0,533.0){\rule[-0.200pt]{0.400pt}{2.409pt}}
\put(544.0,113.0){\rule[-0.200pt]{0.400pt}{2.409pt}}
\put(544.0,533.0){\rule[-0.200pt]{0.400pt}{2.409pt}}
\put(555.0,113.0){\rule[-0.200pt]{0.400pt}{2.409pt}}
\put(555.0,533.0){\rule[-0.200pt]{0.400pt}{2.409pt}}
\put(564.0,113.0){\rule[-0.200pt]{0.400pt}{2.409pt}}
\put(564.0,533.0){\rule[-0.200pt]{0.400pt}{2.409pt}}
\put(572.0,113.0){\rule[-0.200pt]{0.400pt}{2.409pt}}
\put(572.0,533.0){\rule[-0.200pt]{0.400pt}{2.409pt}}
\put(579.0,113.0){\rule[-0.200pt]{0.400pt}{2.409pt}}
\put(579.0,533.0){\rule[-0.200pt]{0.400pt}{2.409pt}}
\put(585.0,113.0){\rule[-0.200pt]{0.400pt}{4.818pt}}
\put(585,68){\makebox(0,0){$1$}}
\put(585.0,523.0){\rule[-0.200pt]{0.400pt}{4.818pt}}
\put(626.0,113.0){\rule[-0.200pt]{0.400pt}{2.409pt}}
\put(626.0,533.0){\rule[-0.200pt]{0.400pt}{2.409pt}}
\put(650.0,113.0){\rule[-0.200pt]{0.400pt}{2.409pt}}
\put(650.0,533.0){\rule[-0.200pt]{0.400pt}{2.409pt}}
\put(666.0,113.0){\rule[-0.200pt]{0.400pt}{2.409pt}}
\put(666.0,533.0){\rule[-0.200pt]{0.400pt}{2.409pt}}
\put(680.0,113.0){\rule[-0.200pt]{0.400pt}{2.409pt}}
\put(680.0,533.0){\rule[-0.200pt]{0.400pt}{2.409pt}}
\put(690.0,113.0){\rule[-0.200pt]{0.400pt}{2.409pt}}
\put(690.0,533.0){\rule[-0.200pt]{0.400pt}{2.409pt}}
\put(699.0,113.0){\rule[-0.200pt]{0.400pt}{2.409pt}}
\put(699.0,533.0){\rule[-0.200pt]{0.400pt}{2.409pt}}
\put(707.0,113.0){\rule[-0.200pt]{0.400pt}{2.409pt}}
\put(707.0,533.0){\rule[-0.200pt]{0.400pt}{2.409pt}}
\put(714.0,113.0){\rule[-0.200pt]{0.400pt}{2.409pt}}
\put(714.0,533.0){\rule[-0.200pt]{0.400pt}{2.409pt}}
\put(720.0,113.0){\rule[-0.200pt]{0.400pt}{4.818pt}}
\put(720,68){\makebox(0,0){$10$}}
\put(720.0,523.0){\rule[-0.200pt]{0.400pt}{4.818pt}}
\put(761.0,113.0){\rule[-0.200pt]{0.400pt}{2.409pt}}
\put(761.0,533.0){\rule[-0.200pt]{0.400pt}{2.409pt}}
\put(220.0,113.0){\rule[-0.200pt]{130.327pt}{0.400pt}}
\put(761.0,113.0){\rule[-0.200pt]{0.400pt}{103.587pt}}
\put(220.0,543.0){\rule[-0.200pt]{130.327pt}{0.400pt}}
\put(133,373){\makebox(0,0){$\alpha(L)$}}
\put(490,23){\makebox(0,0){$L/\xi $}}
\put(355,487){\makebox(0,0)[l]{\small insulator}}
\put(355,282){\makebox(0,0)[l]{\small metal}}
\put(220.0,113.0){\rule[-0.200pt]{0.400pt}{103.587pt}}
\sbox{\plotpoint}{\rule[-0.400pt]{0.800pt}{0.800pt}}%
\put(636,293){\raisebox{-.8pt}{\makebox(0,0){$\Diamond$}}}
\put(653,290){\raisebox{-.8pt}{\makebox(0,0){$\Diamond$}}}
\put(676,273){\raisebox{-.8pt}{\makebox(0,0){$\Diamond$}}}
\put(693,245){\raisebox{-.8pt}{\makebox(0,0){$\Diamond$}}}
\put(706,229){\raisebox{-.8pt}{\makebox(0,0){$\Diamond$}}}
\put(717,217){\raisebox{-.8pt}{\makebox(0,0){$\Diamond$}}}
\put(726,218){\raisebox{-.8pt}{\makebox(0,0){$\Diamond$}}}
\sbox{\plotpoint}{\rule[-0.500pt]{1.000pt}{1.000pt}}%
\put(556,333){\makebox(0,0){$+$}}
\put(573,331){\makebox(0,0){$+$}}
\put(596,318){\makebox(0,0){$+$}}
\put(613,308){\makebox(0,0){$+$}}
\put(626,302){\makebox(0,0){$+$}}
\put(637,299){\makebox(0,0){$+$}}
\put(646,285){\makebox(0,0){$+$}}
\sbox{\plotpoint}{\rule[-0.400pt]{0.800pt}{0.800pt}}%
\put(504,356){\raisebox{-.8pt}{\makebox(0,0){$\Box$}}}
\put(520,354){\raisebox{-.8pt}{\makebox(0,0){$\Box$}}}
\put(544,345){\raisebox{-.8pt}{\makebox(0,0){$\Box$}}}
\put(561,337){\raisebox{-.8pt}{\makebox(0,0){$\Box$}}}
\put(574,329){\raisebox{-.8pt}{\makebox(0,0){$\Box$}}}
\put(585,329){\raisebox{-.8pt}{\makebox(0,0){$\Box$}}}
\put(594,325){\raisebox{-.8pt}{\makebox(0,0){$\Box$}}}
\sbox{\plotpoint}{\rule[-0.500pt]{1.000pt}{1.000pt}}%
\put(379,369){\makebox(0,0){$\times$}}
\put(396,371){\makebox(0,0){$\times$}}
\put(420,372){\makebox(0,0){$\times$}}
\put(437,368){\makebox(0,0){$\times$}}
\put(450,367){\makebox(0,0){$\times$}}
\put(460,365){\makebox(0,0){$\times$}}
\put(470,367){\makebox(0,0){$\times$}}
\sbox{\plotpoint}{\rule[-0.400pt]{0.800pt}{0.800pt}}%
\put(277,377){\makebox(0,0){$\triangle$}}
\put(294,383){\makebox(0,0){$\triangle$}}
\put(318,383){\makebox(0,0){$\triangle$}}
\put(335,384){\makebox(0,0){$\triangle$}}
\put(348,383){\makebox(0,0){$\triangle$}}
\put(358,384){\makebox(0,0){$\triangle$}}
\put(367,385){\makebox(0,0){$\triangle$}}
\sbox{\plotpoint}{\rule[-0.500pt]{1.000pt}{1.000pt}}%
\put(406,390){\makebox(0,0){$\star$}}
\put(423,392){\makebox(0,0){$\star$}}
\put(447,393){\makebox(0,0){$\star$}}
\put(464,397){\makebox(0,0){$\star$}}
\put(477,400){\makebox(0,0){$\star$}}
\put(488,401){\makebox(0,0){$\star$}}
\put(497,401){\makebox(0,0){$\star$}}
\sbox{\plotpoint}{\rule[-0.400pt]{0.800pt}{0.800pt}}%
\put(488,398){\circle{18}}
\put(505,405){\circle{18}}
\put(529,412){\circle{18}}
\put(546,416){\circle{18}}
\put(559,423){\circle{18}}
\put(570,426){\circle{18}}
\put(579,430){\circle{18}}
\sbox{\plotpoint}{\rule[-0.500pt]{1.000pt}{1.000pt}}%
\put(558,421){\circle*{18}}
\put(575,427){\circle*{18}}
\put(599,439){\circle*{18}}
\put(616,447){\circle*{18}}
\put(629,454){\circle*{18}}
\put(640,456){\circle*{18}}
\put(649,461){\circle*{18}}
\end{picture}
}
\end{picture}
\caption[]{One--parameter scaling behavior of the level statistics.
The quantity $\alpha$ as a function of $L/\xi(W)$.}
\label{fig3}
\end{minipage}
\end{figure}
We calculated also
the critical exponent of the localization length $\nu$. 
Using the singularity of $\xi$  near the transition point 
$\xi(W) \propto \mid W-W_{c}\mid ^{-\nu}$ 
one can expand 
the relation (\ref{scaling}) to a power series. Taking into account only
the linear term, 
$\alpha(W,L) = \alpha_{c}(L) + {\rm const}\,L^{1/\nu}\,(W-W_{c})$, and 
applying the $\chi^2$--criterion to fit the data plotted in  Fig.~\ref{fig2}
we found that $\nu=1.45\pm 0.1$.

Another important quantity which is used to describe the level 
statistics is the 
magnitude of fluctuations of the number 
of energy levels 
$\delta N(E)$ in a given energy interval $E$.  
The variance  $\langle [\delta N(E)]^2 \rangle$ 
characterizes the ``stiffness'' of the spectrum.
From the statistical viewpoint it is reasonable to investigate
the dependence of the variance 
$\langle [\delta N(E)]^2 \rangle$ 
on the average number of levels $\langle N(E)\rangle $ in the vicinity
of the MIT. 
Here $\langle...\rangle$ denotes the
averaging over the random configurations with the same disorder $W$.
In the metallic regime the variance 
is known to be defined by the Dyson formula,
$\langle \delta^2 N \rangle_{\rm M} 
= 2/\pi^2\,(\ln \langle N \rangle +  C)$, 
where $C \approx 2.18$, provided that $\langle N \rangle\gg 1$~\cite{RMT}.
In the strongly localized regime the levels are not correlated, hence
$\langle \delta^2 N \rangle_{\rm I} = \langle N \rangle$, 
that is much larger than $\langle \delta^2 N \rangle_{\rm M}$. 
Decreasing disorder suppresses
the fluctuations $\delta N$, so that the variance changes 
from $\langle [\delta N]^2 \rangle_{\rm I}$ 
to $\langle [\delta N]^2 \rangle_{\rm M}$~\cite{Altshuler88}.
Exactly at the transition the variance behaves linearly with 
the average level number, 
$\langle \delta^2 N \rangle_{c} = 
\kappa \langle N \rangle$~\cite{Altshuler88,Thisproc},
as in the insulating regime. 
However the numerical factor $\kappa$ is less than unity.

\begin{figure}[tb]
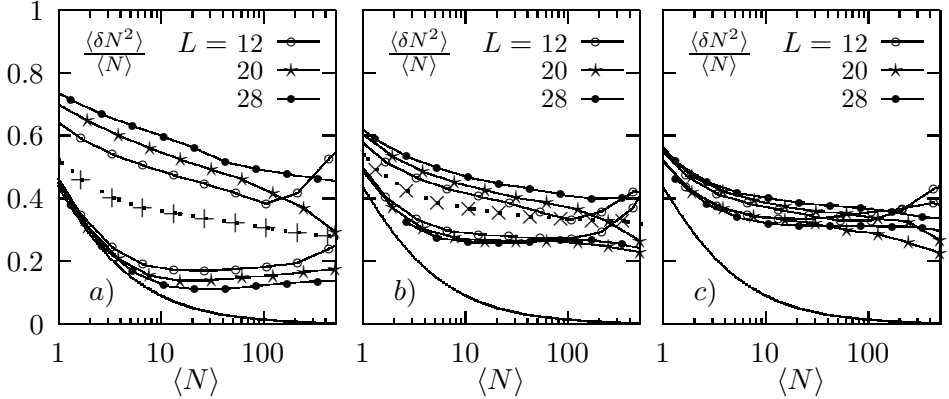

%\vspace{5.2cm}
\unitlength1cm
\begin{minipage}[t]{3.6cm}

}
\end{picture}
\label{fig4c}
\end{minipage}
\caption[]{The variance $\langle [\delta N(E)]^2 \rangle$ as a
function of $\langle N \rangle$ for various sizes $L$ at 
disorder $W$: a) 12 and 20; b) 15 and 18; c) 16 and 17 
for lower and upper set, respectively. 
Critical variance at $W_c=16.5$ 
is shown for different $L$: a) 24~(+); b) 28~($\times$).
Solid line corresponds to the Dyson law.
}
\label{fig4}
\end{figure}

We calculated the dependence of the ratio  
$\langle [\delta N(E)]^2 \rangle/\langle N(E) \rangle$
on the average number of levels within a given interval $E$ for different
lattice sizes $L$ at the disorder $W$ varying
from 12 to 20, as shown in Fig.~\ref{fig4}. 
It is obvious that near the MIT this ratio exhibits the critical behavior.
In approaching the transition, $W=W_{c}$, it becomes size--invariant
(Fig.~4c). Such a behavior is closely related to the
universality of the critical level spacing distribution, as mentioned above. 
In addition, one can observe  that at the MIT 
the relative fluctuations
of the level number
$\langle [\delta N(E)]^2 \rangle/\langle N(E) \rangle$ 
decreases with the 
energy, when $\langle N \rangle \equiv E/\Delta \sim 1$, 
and then varies very weakly
over two orders of magnitude of $\langle N \rangle$,
tending to a constant value  
$\kappa \approx 0.32$.
For 
$E>200\Delta$ the numerical errors becomes larger
due to the finite number of realizations.
Our results are consistent with the suggestion about the proportionality 
between the variance and the average level number~\cite{Altshuler88},
but deviate from the power law proposed recently in~\cite{Kravtsov94}.

Determining the ratio 
$\langle \delta N(E)^2 \rangle/\langle N(E) \rangle$
as the function 
$F(E,L,W)$, one can analyze the scaling properties 
similarly to those of $P(s)$. For example,
the disorder dependence 
for the given energy width $E=20\Delta$ is shown in Fig.~\ref{fig5}. 
Near the critical point, $|W-W_{c}|<1$, 
the function can be linearized 
$F(E,L,W) = \kappa + A\,L^{1/\nu}\,(W-W_{c})$. 
The factor $A$ depends on the energy $E$, in contrast to that for
$\alpha$, 
whose critical behavior is not sensitive to the choice of $s$.
In order to obtain the one--parameter scaling law 
one should take in account the typical volume 
per one electron state lying in the interval $E$,
$L_{o}^3=L^3\,(\Delta/E)$, instead of the total volume $L^3$, i.e.
$F-\kappa \propto (L_{o}/\xi)^{1/\nu}$. 
Fig.~6 shows $\langle \delta N^2 \rangle /\langle N \rangle$ 
vs. $L_{o}/\xi(W)$, where $\xi(W)$ is taken from the 
analysis of $\alpha$.
All data belong to the common two--branch curve regardless
to $L$, $W$, and chosen $\langle N \rangle$. 
Indeed, the fluctuations $\delta N$ in the interval $E$ are mainly
defined by the states confined within the correlation volume $\xi$, 
which are separated by the energy $\Delta_{\xi}=\Delta\,(L/\xi)^3$.
The number of those states is $N_{\xi}= E/\Delta_{\xi}$. Therefore
$F = \kappa + N_{\xi}^{-1/3\nu}\,{\rm sign} (W-W_{c})$.
\begin{figure}[tb]
%\vspace{5.2cm}
\unitlength1cm
\begin{minipage}[t]{6.0cm}
\begin{picture}(6.0,5.)
\put(-0.8,-0.2){
%%\input{plot8}
%%-------------------------------------------------
% GNUPLOT: LaTeX picture
\setlength{\unitlength}{0.240900pt}
\ifx\plotpoint\undefined\newsavebox{\plotpoint}\fi
\sbox{\plotpoint}{\rule[-0.200pt]{0.400pt}{0.400pt}}%
\begin{picture}(825,629)(0,0)
\font\gnuplot=cmr10 at 10pt
\gnuplot
\sbox{\plotpoint}{\rule[-0.200pt]{0.400pt}{0.400pt}}%
\put(220.0,113.0){\rule[-0.200pt]{4.818pt}{0.400pt}}
\put(198,113){\makebox(0,0)[r]{0.1}}
\put(741.0,113.0){\rule[-0.200pt]{4.818pt}{0.400pt}}
\put(220.0,212.0){\rule[-0.200pt]{4.818pt}{0.400pt}}
\put(741.0,212.0){\rule[-0.200pt]{4.818pt}{0.400pt}}
\put(220.0,310.0){\rule[-0.200pt]{4.818pt}{0.400pt}}
\put(198,310){\makebox(0,0)[r]{0.3}}
\put(741.0,310.0){\rule[-0.200pt]{4.818pt}{0.400pt}}
\put(220.0,409.0){\rule[-0.200pt]{4.818pt}{0.400pt}}
\put(741.0,409.0){\rule[-0.200pt]{4.818pt}{0.400pt}}
\put(220.0,507.0){\rule[-0.200pt]{4.818pt}{0.400pt}}
\put(198,507){\makebox(0,0)[r]{0.5}}
\put(741.0,507.0){\rule[-0.200pt]{4.818pt}{0.400pt}}
\put(220.0,606.0){\rule[-0.200pt]{4.818pt}{0.400pt}}
\put(741.0,606.0){\rule[-0.200pt]{4.818pt}{0.400pt}}
\put(274.0,113.0){\rule[-0.200pt]{0.400pt}{4.818pt}}
\put(274,68){\makebox(0,0){12}}
\put(274.0,586.0){\rule[-0.200pt]{0.400pt}{4.818pt}}
\put(382.0,113.0){\rule[-0.200pt]{0.400pt}{4.818pt}}
\put(382,68){\makebox(0,0){14}}
\put(382.0,586.0){\rule[-0.200pt]{0.400pt}{4.818pt}}
\put(491.0,113.0){\rule[-0.200pt]{0.400pt}{4.818pt}}
\put(491,68){\makebox(0,0){16}}
\put(491.0,586.0){\rule[-0.200pt]{0.400pt}{4.818pt}}
\put(599.0,113.0){\rule[-0.200pt]{0.400pt}{4.818pt}}
\put(599,68){\makebox(0,0){18}}
\put(599.0,586.0){\rule[-0.200pt]{0.400pt}{4.818pt}}
\put(707.0,113.0){\rule[-0.200pt]{0.400pt}{4.818pt}}
\put(707,68){\makebox(0,0){20}}
\put(707.0,586.0){\rule[-0.200pt]{0.400pt}{4.818pt}}
\put(220.0,113.0){\rule[-0.200pt]{130.327pt}{0.400pt}}
\put(761.0,113.0){\rule[-0.200pt]{0.400pt}{118.764pt}}
\put(220.0,606.0){\rule[-0.200pt]{130.327pt}{0.400pt}}
\put(133,404){\makebox(0,0){$\frac{\langle \delta N^2 \rangle}{\langle N \rangle}$}}
\put(490,23){\makebox(0,0){$W$}}
\put(220.0,113.0){\rule[-0.200pt]{0.400pt}{118.764pt}}
\put(436,557){\makebox(0,0)[r]{$L=$ 12}}
\put(458.0,557.0){\rule[-0.200pt]{15.899pt}{0.400pt}}
\put(274,183){\usebox{\plotpoint}}
\multiput(274.00,183.58)(0.795,0.499){133}{\rule{0.735pt}{0.120pt}}
\multiput(274.00,182.17)(106.474,68.000){2}{\rule{0.368pt}{0.400pt}}
\multiput(382.00,251.58)(0.643,0.498){81}{\rule{0.614pt}{0.120pt}}
\multiput(382.00,250.17)(52.725,42.000){2}{\rule{0.307pt}{0.400pt}}
\multiput(436.00,293.58)(0.561,0.498){95}{\rule{0.549pt}{0.120pt}}
\multiput(436.00,292.17)(53.861,49.000){2}{\rule{0.274pt}{0.400pt}}
\multiput(491.00,342.61)(5.820,0.447){3}{\rule{3.700pt}{0.108pt}}
\multiput(491.00,341.17)(19.320,3.000){2}{\rule{1.850pt}{0.400pt}}
\multiput(518.00,345.58)(0.713,0.495){35}{\rule{0.668pt}{0.119pt}}
\multiput(518.00,344.17)(25.613,19.000){2}{\rule{0.334pt}{0.400pt}}
\multiput(545.00,364.58)(0.676,0.498){77}{\rule{0.640pt}{0.120pt}}
\multiput(545.00,363.17)(52.672,40.000){2}{\rule{0.320pt}{0.400pt}}
\multiput(599.00,404.58)(0.845,0.499){125}{\rule{0.775pt}{0.120pt}}
\multiput(599.00,403.17)(106.391,64.000){2}{\rule{0.388pt}{0.400pt}}
\put(480,557){\circle{12}}
\put(274,183){\circle{12}}
\put(382,251){\circle{12}}
\put(436,293){\circle{12}}
\put(491,342){\circle{12}}
\put(518,345){\circle{12}}
\put(545,364){\circle{12}}
\put(599,404){\circle{12}}
\put(707,468){\circle{12}}
\put(436,512){\makebox(0,0)[r]{20}}
\put(458.0,512.0){\rule[-0.200pt]{15.899pt}{0.400pt}}
\put(274,151){\usebox{\plotpoint}}
\multiput(274.00,151.58)(0.659,0.499){161}{\rule{0.627pt}{0.120pt}}
\multiput(274.00,150.17)(106.699,82.000){2}{\rule{0.313pt}{0.400pt}}
\multiput(382.00,233.58)(0.628,0.498){83}{\rule{0.602pt}{0.120pt}}
\multiput(382.00,232.17)(52.750,43.000){2}{\rule{0.301pt}{0.400pt}}
\multiput(436.58,276.00)(0.499,0.554){107}{\rule{0.120pt}{0.544pt}}
\multiput(435.17,276.00)(55.000,59.872){2}{\rule{0.400pt}{0.272pt}}
\multiput(491.00,337.58)(0.587,0.496){43}{\rule{0.570pt}{0.120pt}}
\multiput(491.00,336.17)(25.818,23.000){2}{\rule{0.285pt}{0.400pt}}
\multiput(518.00,360.58)(0.713,0.495){35}{\rule{0.668pt}{0.119pt}}
\multiput(518.00,359.17)(25.613,19.000){2}{\rule{0.334pt}{0.400pt}}
\multiput(545.00,379.58)(0.539,0.498){97}{\rule{0.532pt}{0.120pt}}
\multiput(545.00,378.17)(52.896,50.000){2}{\rule{0.266pt}{0.400pt}}
\multiput(599.00,429.58)(0.551,0.499){193}{\rule{0.541pt}{0.120pt}}
\multiput(599.00,428.17)(106.878,98.000){2}{\rule{0.270pt}{0.400pt}}
\put(480,512){\makebox(0,0){$\star$}}
\put(274,151){\makebox(0,0){$\star$}}
\put(382,233){\makebox(0,0){$\star$}}
\put(436,276){\makebox(0,0){$\star$}}
\put(491,337){\makebox(0,0){$\star$}}
\put(518,360){\makebox(0,0){$\star$}}
\put(545,379){\makebox(0,0){$\star$}}
\put(599,429){\makebox(0,0){$\star$}}
\put(707,527){\makebox(0,0){$\star$}}
\put(436,467){\makebox(0,0)[r]{28}}
\put(458.0,467.0){\rule[-0.200pt]{15.899pt}{0.400pt}}
\put(274,124){\usebox{\plotpoint}}
\multiput(274.00,124.58)(0.628,0.499){169}{\rule{0.602pt}{0.120pt}}
\multiput(274.00,123.17)(106.750,86.000){2}{\rule{0.301pt}{0.400pt}}
\multiput(382.58,210.00)(0.498,0.546){105}{\rule{0.120pt}{0.537pt}}
\multiput(381.17,210.00)(54.000,57.885){2}{\rule{0.400pt}{0.269pt}}
\multiput(436.00,269.58)(0.499,0.499){107}{\rule{0.500pt}{0.120pt}}
\multiput(436.00,268.17)(53.962,55.000){2}{\rule{0.250pt}{0.400pt}}
\multiput(491.58,324.00)(0.497,0.724){51}{\rule{0.120pt}{0.678pt}}
\multiput(490.17,324.00)(27.000,37.593){2}{\rule{0.400pt}{0.339pt}}
\multiput(518.58,363.00)(0.497,0.611){51}{\rule{0.120pt}{0.589pt}}
\multiput(517.17,363.00)(27.000,31.778){2}{\rule{0.400pt}{0.294pt}}
\multiput(545.58,396.00)(0.498,0.574){105}{\rule{0.120pt}{0.559pt}}
\multiput(544.17,396.00)(54.000,60.839){2}{\rule{0.400pt}{0.280pt}}
\multiput(599.58,458.00)(0.499,0.504){213}{\rule{0.120pt}{0.504pt}}
\multiput(598.17,458.00)(108.000,107.955){2}{\rule{0.400pt}{0.252pt}}
\put(480,467){\circle*{12}}
\put(274,124){\circle*{12}}
\put(382,210){\circle*{12}}
\put(436,269){\circle*{12}}
\put(491,324){\circle*{12}}
\put(518,363){\circle*{12}}
\put(545,396){\circle*{12}}
\put(599,458){\circle*{12}}
\put(707,567){\circle*{12}}
\end{picture}
}
\end{picture}
\caption[]{Disorder dependence of the ratio 
$\langle \delta N^2 \rangle /\langle N \rangle$ 
at the given number of levels 
$\langle N \rangle \equiv E/\Delta = 20$  
for various sizes $L$. 
}
\label{fig5}
\end{minipage}
\hfill
\begin{minipage}[t]{6cm}
\begin{picture}(6.0,5.)
\put(-0.7,-0.2){
%%\input{plot9}
%%------------------------------------------------------
% GNUPLOT: LaTeX picture
\setlength{\unitlength}{0.240900pt}
\ifx\plotpoint\undefined\newsavebox{\plotpoint}\fi
\sbox{\plotpoint}{\rule[-0.200pt]{0.400pt}{0.400pt}}%
\begin{picture}(825,629)(0,0)
\font\gnuplot=cmr10 at 10pt
\gnuplot
\sbox{\plotpoint}{\rule[-0.200pt]{0.400pt}{0.400pt}}%
\put(220.0,360.0){\rule[-0.200pt]{130.327pt}{0.400pt}}
\put(220.0,113.0){\rule[-0.200pt]{4.818pt}{0.400pt}}
\put(198,113){\makebox(0,0)[r]{-0.3}}
\put(741.0,113.0){\rule[-0.200pt]{4.818pt}{0.400pt}}
\put(220.0,195.0){\rule[-0.200pt]{4.818pt}{0.400pt}}
\put(741.0,195.0){\rule[-0.200pt]{4.818pt}{0.400pt}}
\put(220.0,277.0){\rule[-0.200pt]{4.818pt}{0.400pt}}
\put(741.0,277.0){\rule[-0.200pt]{4.818pt}{0.400pt}}
\put(220.0,360.0){\rule[-0.200pt]{4.818pt}{0.400pt}}
\put(198,360){\makebox(0,0)[r]{0}}
\put(741.0,360.0){\rule[-0.200pt]{4.818pt}{0.400pt}}
\put(220.0,442.0){\rule[-0.200pt]{4.818pt}{0.400pt}}
\put(741.0,442.0){\rule[-0.200pt]{4.818pt}{0.400pt}}
\put(220.0,524.0){\rule[-0.200pt]{4.818pt}{0.400pt}}
\put(741.0,524.0){\rule[-0.200pt]{4.818pt}{0.400pt}}
\put(220.0,606.0){\rule[-0.200pt]{4.818pt}{0.400pt}}
\put(198,606){\makebox(0,0)[r]{0.3}}
\put(741.0,606.0){\rule[-0.200pt]{4.818pt}{0.400pt}}
\put(220.0,113.0){\rule[-0.200pt]{0.400pt}{2.409pt}}
\put(220.0,596.0){\rule[-0.200pt]{0.400pt}{2.409pt}}
\put(230.0,113.0){\rule[-0.200pt]{0.400pt}{2.409pt}}
\put(230.0,596.0){\rule[-0.200pt]{0.400pt}{2.409pt}}
\put(239.0,113.0){\rule[-0.200pt]{0.400pt}{2.409pt}}
\put(239.0,596.0){\rule[-0.200pt]{0.400pt}{2.409pt}}
\put(247.0,113.0){\rule[-0.200pt]{0.400pt}{4.818pt}}
\put(247,68){\makebox(0,0){$0.01$}}
\put(247.0,586.0){\rule[-0.200pt]{0.400pt}{4.818pt}}
\put(298.0,113.0){\rule[-0.200pt]{0.400pt}{2.409pt}}
\put(298.0,596.0){\rule[-0.200pt]{0.400pt}{2.409pt}}
\put(328.0,113.0){\rule[-0.200pt]{0.400pt}{2.409pt}}
\put(328.0,596.0){\rule[-0.200pt]{0.400pt}{2.409pt}}
\put(350.0,113.0){\rule[-0.200pt]{0.400pt}{2.409pt}}
\put(350.0,596.0){\rule[-0.200pt]{0.400pt}{2.409pt}}
\put(366.0,113.0){\rule[-0.200pt]{0.400pt}{2.409pt}}
\put(366.0,596.0){\rule[-0.200pt]{0.400pt}{2.409pt}}
\put(380.0,113.0){\rule[-0.200pt]{0.400pt}{2.409pt}}
\put(380.0,596.0){\rule[-0.200pt]{0.400pt}{2.409pt}}
\put(391.0,113.0){\rule[-0.200pt]{0.400pt}{2.409pt}}
\put(391.0,596.0){\rule[-0.200pt]{0.400pt}{2.409pt}}
\put(401.0,113.0){\rule[-0.200pt]{0.400pt}{2.409pt}}
\put(401.0,596.0){\rule[-0.200pt]{0.400pt}{2.409pt}}
\put(410.0,113.0){\rule[-0.200pt]{0.400pt}{2.409pt}}
\put(410.0,596.0){\rule[-0.200pt]{0.400pt}{2.409pt}}
\put(418.0,113.0){\rule[-0.200pt]{0.400pt}{4.818pt}}
\put(418,68){\makebox(0,0){$0.1$}}
\put(418.0,586.0){\rule[-0.200pt]{0.400pt}{4.818pt}}
\put(470.0,113.0){\rule[-0.200pt]{0.400pt}{2.409pt}}
\put(470.0,596.0){\rule[-0.200pt]{0.400pt}{2.409pt}}
\put(500.0,113.0){\rule[-0.200pt]{0.400pt}{2.409pt}}
\put(500.0,596.0){\rule[-0.200pt]{0.400pt}{2.409pt}}
\put(521.0,113.0){\rule[-0.200pt]{0.400pt}{2.409pt}}
\put(521.0,596.0){\rule[-0.200pt]{0.400pt}{2.409pt}}
\put(538.0,113.0){\rule[-0.200pt]{0.400pt}{2.409pt}}
\put(538.0,596.0){\rule[-0.200pt]{0.400pt}{2.409pt}}
\put(551.0,113.0){\rule[-0.200pt]{0.400pt}{2.409pt}}
\put(551.0,596.0){\rule[-0.200pt]{0.400pt}{2.409pt}}
\put(563.0,113.0){\rule[-0.200pt]{0.400pt}{2.409pt}}
\put(563.0,596.0){\rule[-0.200pt]{0.400pt}{2.409pt}}
\put(573.0,113.0){\rule[-0.200pt]{0.400pt}{2.409pt}}
\put(573.0,596.0){\rule[-0.200pt]{0.400pt}{2.409pt}}
\put(582.0,113.0){\rule[-0.200pt]{0.400pt}{2.409pt}}
\put(582.0,596.0){\rule[-0.200pt]{0.400pt}{2.409pt}}
\put(590.0,113.0){\rule[-0.200pt]{0.400pt}{4.818pt}}
\put(590,68){\makebox(0,0){$1$}}
\put(590.0,586.0){\rule[-0.200pt]{0.400pt}{4.818pt}}
\put(641.0,113.0){\rule[-0.200pt]{0.400pt}{2.409pt}}
\put(641.0,596.0){\rule[-0.200pt]{0.400pt}{2.409pt}}
\put(671.0,113.0){\rule[-0.200pt]{0.400pt}{2.409pt}}
\put(671.0,596.0){\rule[-0.200pt]{0.400pt}{2.409pt}}
\put(693.0,113.0){\rule[-0.200pt]{0.400pt}{2.409pt}}
\put(693.0,596.0){\rule[-0.200pt]{0.400pt}{2.409pt}}
\put(709.0,113.0){\rule[-0.200pt]{0.400pt}{2.409pt}}
\put(709.0,596.0){\rule[-0.200pt]{0.400pt}{2.409pt}}
\put(723.0,113.0){\rule[-0.200pt]{0.400pt}{2.409pt}}
\put(723.0,596.0){\rule[-0.200pt]{0.400pt}{2.409pt}}
\put(734.0,113.0){\rule[-0.200pt]{0.400pt}{2.409pt}}
\put(734.0,596.0){\rule[-0.200pt]{0.400pt}{2.409pt}}
\put(744.0,113.0){\rule[-0.200pt]{0.400pt}{2.409pt}}
\put(744.0,596.0){\rule[-0.200pt]{0.400pt}{2.409pt}}
\put(753.0,113.0){\rule[-0.200pt]{0.400pt}{2.409pt}}
\put(753.0,596.0){\rule[-0.200pt]{0.400pt}{2.409pt}}
\put(761.0,113.0){\rule[-0.200pt]{0.400pt}{4.818pt}}
\put(761,68){\makebox(0,0){$10$}}
\put(761.0,586.0){\rule[-0.200pt]{0.400pt}{4.818pt}}
\put(220.0,113.0){\rule[-0.200pt]{130.327pt}{0.400pt}}
\put(761.0,113.0){\rule[-0.200pt]{0.400pt}{118.764pt}}
\put(220.0,606.0){\rule[-0.200pt]{130.327pt}{0.400pt}}
\put(111,494){\makebox(0,0){$\frac{\langle \delta N^2\rangle}{\langle N\rangle}-\kappa$}}
\put(490,23){\makebox(0,0){$L_{o}/\xi$}}
\put(220.0,113.0){\rule[-0.200pt]{0.400pt}{118.764pt}}
\put(401,253){\makebox(0,0)[r]{$L=$ 12}}
\put(445,253){\circle{12}}
\put(606,240){\circle{12}}
\put(624,226){\circle{12}}
\put(641,214){\circle{12}}
\put(505,292){\circle{12}}
\put(522,281){\circle{12}}
\put(539,271){\circle{12}}
\put(439,322){\circle{12}}
\put(456,315){\circle{12}}
\put(473,307){\circle{12}}
\put(281,380){\circle{12}}
\put(298,360){\circle{12}}
\put(315,349){\circle{12}}
\put(239,346){\circle{12}}
\put(257,350){\circle{12}}
\put(274,353){\circle{12}}
\put(351,366){\circle{12}}
\put(368,370){\circle{12}}
\put(385,373){\circle{12}}
\put(437,390){\circle{12}}
\put(455,401){\circle{12}}
\put(472,407){\circle{12}}
\put(516,439){\circle{12}}
\put(534,453){\circle{12}}
\put(551,462){\circle{12}}
\put(401,208){\makebox(0,0)[r]{20}}
\put(445,208){\makebox(0,0){$\star$}}
\put(641,220){\makebox(0,0){$\star$}}
\put(658,203){\makebox(0,0){$\star$}}
\put(675,189){\makebox(0,0){$\star$}}
\put(539,284){\makebox(0,0){$\star$}}
\put(557,270){\makebox(0,0){$\star$}}
\put(574,255){\makebox(0,0){$\star$}}
\put(473,316){\makebox(0,0){$\star$}}
\put(490,305){\makebox(0,0){$\star$}}
\put(508,295){\makebox(0,0){$\star$}}
\put(315,345){\makebox(0,0){$\star$}}
\put(333,352){\makebox(0,0){$\star$}}
\put(350,349){\makebox(0,0){$\star$}}
\put(274,376){\makebox(0,0){$\star$}}
\put(291,372){\makebox(0,0){$\star$}}
\put(308,366){\makebox(0,0){$\star$}}
\put(385,388){\makebox(0,0){$\star$}}
\put(402,384){\makebox(0,0){$\star$}}
\put(420,385){\makebox(0,0){$\star$}}
\put(472,417){\makebox(0,0){$\star$}}
\put(489,420){\makebox(0,0){$\star$}}
\put(506,427){\makebox(0,0){$\star$}}
\put(551,478){\makebox(0,0){$\star$}}
\put(568,492){\makebox(0,0){$\star$}}
\put(585,508){\makebox(0,0){$\star$}}
\put(401,163){\makebox(0,0)[r]{28}}
\put(445,163){\circle*{12}}
\put(658,202){\circle*{12}}
\put(675,185){\circle*{12}}
\put(693,172){\circle*{12}}
\put(710,170){\circle*{12}}
\put(557,280){\circle*{12}}
\put(574,262){\circle*{12}}
\put(591,244){\circle*{12}}
\put(608,232){\circle*{12}}
\put(491,324){\circle*{12}}
\put(508,308){\circle*{12}}
\put(525,294){\circle*{12}}
\put(542,283){\circle*{12}}
\put(333,360){\circle*{12}}
\put(350,350){\circle*{12}}
\put(367,339){\circle*{12}}
\put(384,332){\circle*{12}}
\put(291,380){\circle*{12}}
\put(308,375){\circle*{12}}
\put(325,372){\circle*{12}}
\put(343,370){\circle*{12}}
\put(402,408){\circle*{12}}
\put(420,403){\circle*{12}}
\put(437,399){\circle*{12}}
\put(454,397){\circle*{12}}
\put(489,444){\circle*{12}}
\put(506,448){\circle*{12}}
\put(523,450){\circle*{12}}
\put(541,453){\circle*{12}}
\put(568,510){\circle*{12}}
\put(585,518){\circle*{12}}
\put(602,543){\circle*{12}}
\put(620,561){\circle*{12}}
\end{picture}
}
\end{picture}
\caption[]{ 
$\langle \delta N^2 \rangle /\langle N \rangle -\kappa$ as a function of 
$L_{o}/\xi(W)$ for different $L$ and various $\langle N \rangle$ from 
10 to 100.}
\label{fig6}
\end{minipage}
\end{figure}

In conclusion, we have used the level spacing distribution $P(s)$ as a 
scaling variable in order to detect the critical behavior at the 
disorder--induced metal--insulator transition. 
The results for the critical exponent $\nu $ 
and the critical disorder $W_{c}$, 
which were obtained by 
numerically diagonalizing the Anderson Hamiltonian for up to $28^{3}$
lattice sizes, are consistent 
with those obtained earlier by using
completely different approaches \cite{Kramer94}. 
Our calculations showed that the universality of the level 
statistics at the transition is revealed
not only in the form of $P(s)$ but also in the dependence of the variance
of the number of electron states 
in an energy interval of a given width
on their mean number $\langle N \rangle$. 
We  also analyzed the scaling behavior of the function
$\langle \delta^2 N \rangle = F(\langle N \rangle)$. 
At the MIT this function is found 
to be $L$--invariant with a leading linear term, 
$\langle \delta^2 N\rangle_{c} = \kappa \langle N \rangle$, 
where $\kappa \approx 0.3$.

\end{document}